\documentclass{elsart3p}

\usepackage{natbib}

   \usepackage{graphicx}
   \usepackage{epstopdf}
  \usepackage{amssymb}

\def\aa{{Astron. Astrophys.}}

\def\apj{ApJ}

\def\nat{Nature}

\def\mnras{{M.N.R.A.S.}}
\def\prl{Phys. Rev. Lett.}                      
\def\reference{\par \noindent \hangafter=1 \hangindent=0.7 true cm}
             \font\sevenrm=cmr7

\def\dover#1#2{\hbox{${{\displaystyle#1 \vphantom{(} }\over{
   \displaystyle #2 \vphantom{(} }}$}}

\begin{document}
\newcommand{\vol}[2]{$\,$\rm #1\rm , #2,}                 
\def\gamsk{\gamma_1}
\def\erg{\varepsilon_\gamma}
\def\dover#1#2{\hbox{${{\displaystyle#1 \vphantom{(} }\over{
   \displaystyle #2 \vphantom{(} }}$}}
\def\thetascatt{\theta_{\hbox{\sevenrm scatt}}}
\def\thetaBone{\Theta_{\hbox{\sevenrm Bn1}}}
\def\thetaBtwo{\Theta_{\hbox{\sevenrm Bn2}}}
{\catcode`\@=11 
  \gdef\SchlangeUnter#1#2{\lower2pt\vbox{\baselineskip 0pt\lineskip0pt 
  \ialign{$\m@th#1\hfil##\hfil$\crcr#2\crcr\sim\crcr}}}} 
\def\gtrsim{\mathrel{\mathpalette\SchlangeUnter>}} 
\def\lesssim{\mathrel{\mathpalette\SchlangeUnter<}} 
\newcommand{\figureoutpdf}[5]{\centerline{}
   \centerline{\hspace{#3in} \includegraphics[width=#2truein]{#1}}
   \vspace{#4truein} \caption{#5} \centerline{} }
\begin{frontmatter}



\title{Gamma-Ray Burst Afterglows}

\author{Bing Zhang}

\address{Department of Physics, University of Nevada, Las Vegas, USA \\
{\tt Email: bzhang@physics.unlv.edu}}

\begin{abstract}

Extended, fading emissions in multi-wavelength are observed following
Gamma-ray bursts (GRBs). Recent broad-band observational campaigns led
by the Swift Observatory reveal rich features of these GRB afterglows.
Here we review the latest observational progress and discuss the
theoretical implications for understanding the central engine, 
composition, and geometric configuration of GRB jets, as well
as their interactions with the ambient medium.

\end{abstract}


\end{frontmatter}
\setlength{\parindent}{.25in}

\section{Introduction}
\label{sec:Introduction}

Gamma-ray bursts (GRBs) are the most violent explosions in the universe.
They are relativistic outflows launched during collapses of 
massive stars or mergers of compact objects. Regardless of the
nature of the explosion, a generic fireball shock model (Rees \&
M\'esz\'aros 1992, 1994; M\'esz\'aros \& Rees 1993, 1997, for reviews see
Piran 1999, 2005; M\'esz\'aros 2002, 2006; Zhang \& M\'esz\'aros 2004)
is found successful to interpret the broad GRB phenomenology. According
to this model, the ejecta is intrinsically intermittent and unsteady, and is
composed of many mini-shells with a wide range of bulk Lorentz factors. 
Internal shocks (Rees \& M\'esz\'aros 1994) are likely developed before 
the global fireball is decelerated by the ambient medium, which are
generally believed to be the emission sites of the observed prompt GRB
emission. The fireball is decelerated at a larger distance after 
sweeping enough interstellar medium whose inertia becomes noticeable, 
and the blastwave gradually enters a self-similar deceleration regime
(Blandford-McKee 1976). Upon deceleration, a pair of shocks
forms. A long-lived forward shock propagating into the ambient medium
gives rise to the long-term broad band afterglow (M\'esz\'aros \& Rees
1997; Sari et al. 1998); and a short-lived reverse shock propagating
into the ejecta itself gives rise to a possible short-term optical/IR
flash and a radio flare (M\'esz\'aros \& Rees 1997, 1999; 
Sari \& Piran 1999a,b). The relativistic ejecta are likely collimated 
(Rhoads 1999; Sari et al. 1999), and the jets may have substantial 
angular structures (Zhang \& M\'esz\'aros 2002; Rossi et al. 2002). 
This general theoretical framework has been successful to interpret 
most of the observational data in the pre-Swift era. 

The successful launch and operation of NASA's broadband (gamma-ray, 
X-ray, UV \& optical) GRB mission Swift (Gehrels et al. 2004) opens 
a brand new era in the GRB study. The prompt slewing capability of
the X-Ray Telescope (XRT, Burrows et al. 2005a) and UV-Optical
Telescope (UVOT, Roming et al. 2005) allows the satellite to swiftly 
catch the very early X-ray and UV/optical signals following 
the GRB prompt emission detected by the Burst Alert Telescope (BAT,
Barthelmy et al. 2005a). The precise localizations made by XRT for
the majority of the bursts make it possible for ground-based 
follow up observations of most bursts. We now have unprecedented 
information about GRB afterglows, which sheds 
light onto many outstanding problems
in the pre-Swift era (Zhang \& M\'esz\'aros 2004 for a summary):
e.g. central engine, composition and geometric configuration of the
GRB fireball, and its interaction with the ambient medium.

\section{A canonical lightcurve of X-ray afterglows}
 \label{sec:X-ray}

\begin{figure}
  \includegraphics[height=.4\textheight,angle=-90]{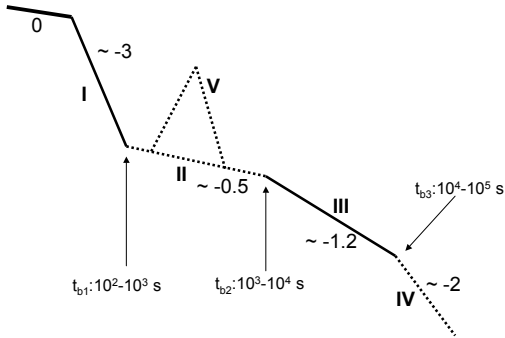}
  \caption{A canonical X-ray afterglow lightcurve revealed by 
Swift XRT observations (from Zhang et al. 2006)}
\label{XRT-lc}
\end{figure}

One of the major discoveries of Swift is the identification of a
canonical X-ray afterglow behavior (Nousek et al. 2006; Zhang et al.
2006; O'Brien et al. 2006, Chincarini et al. 2005; 
see Fig.1). Although different afterglow
lightcurves may vary from one another, they are all composed of 
several of the five components illustrated in Fig.1. 

\begin{itemize}
\item Steep decay phase (I): Typically smoothly connected to the prompt 
emission (Tagliaferri et al. 2005; Barthelmy et al. 2005b), with a 
temporal decay slope $\sim -3$ or steeper (sometimes up to $\sim -10$, e.g.
Vaughan et al. 2006; Cusumano et al. 2006; O'Brien et al. 2006) extending
to $\sim (10^2-10^3)$s. Usually have a different spectral slope from 
the later afterglow phases. 
\item Shallow decay phase (II): Typically with a temporal decay slope
$\sim -0.5$ or flatter extending to $\sim (10^3-10^4)$s, at which a 
temporal break is observed before the normal decay phase (e.g. Campana et al.
2005; De Pasquale et al. 2006). There is no spectral evolution across the
break.
\item Normal decay phase (III): Usually with a decay slope $\sim -1.2$,
and usually follows the predictions of the standard afterglow model
(M\'esz\'aros \& Rees 1997; Sari et al. 1998; Chevalier \& Li 2000).
A systematic test of the afterglow closure-relations (e.g. Table 1 of
Zhang \& M\'esz\'aros 2004) suggests however that a fraction of bursts
do not satisfy any afterglow model (Willingale et al. 2006).
\item Post Jet break phase (IV): Occasionally observed following the 
normal decay phase, typically with a decay slope $\sim -2$, satisfying 
the predictions of the standard jet model (Rhoads 1999; Sari et al. 1999)
or the structured jet model (Rossi et al. 2002; Zhang \& M\'esz\'aros 
2002).
\item X-ray flares (V): Appear in nearly half of GRB afterglows. Sometimes
multiple flares harbor in one GRB. Typically have very steep rising and
decaying slopes (Burrows et al. 2005b; Falcone et al. 2006; Romano et al.
2006) with $\delta t/t \ll 1$. Appear in both long-duration 
(Falcone et al. 2006) and short-duration GRBs (Barthelmy et al. 2005b;
Campana et al. 2006a), and both hard GRBs and soft X-ray flashes
(Romano et al. 2006).
\end{itemize}

Except for the normal decay and the jet-break phases, all the other 
three components were not straightforwardly expected in the pre-Swift
era\footnote{The flare-like signature was seen by Beppo-SAX, but it
was interpreted as the onset of the afterglow (Piro et al. 2005).}.
As of the time of writing, the steep decay phase and X-ray flares
are better understood, while the shallow decay phase is still
a mystery.

\subsection{Steep decay phase: tail of the prompt emission}
 \label{sec:steep}

The generally accepted interpretation of the steep decay phase is the
tail emission due to the so-called ``curvature effect'' (Fenimore et
al. 1996; Kumar \& Panaitescu 2000; Zhang et al. 2006; Panaitescu
et al. 2006a; Dyks et al. 2006). The basic assumption of this
interpretation is that the GRB emission region is disconnected from
the afterglow region (the external shock), and that the emission from
the GRB emission region ceases abruptly. This is consistent with the
conjectures of internal shocks or other internal dissipation mechanisms
(e.g. magnetic fields reconnection, etc). Since it is generally
assumed that the ejecta has a conical geometry, the curvature of the
radiation front causes a propagation delay for high-latitude emission
from the line of sight. Combining with the variation of the Doppler
factor at different latitudes, one gets a simple prediction 
\begin{equation}
\alpha=2+\beta
\label{curvature}
\end{equation} 
for the emission outside the $\Gamma^{-1}$ emission
cone, where the convention $F_\nu \propto t^{-\alpha} \nu^{-\beta}$
is adopted. The salient feature of this interpretation is that 
it could be directly tested since both $\alpha$ and $\beta$ could be
measured directly from the observational data, given that two
complications are treated properly (Zhang et al. 2006): First, 
for internal emissions, every time when the central engine restarts,
the clock should be re-set to zero\footnote{We notice that for 
external shock related emissions, taking the GRB trigger time as
the time zero point is generally required (Lazzati \& Begelman
2006; Kobayashi \& Zhang 2006).}. In the $\log - \log$ lightcurves,
this usually introduces an ``artificial'' very steep decay if the 
GRB trigger time (which is usually taken as $t=0$) significantly 
leads the time zero point ($t_0$) of the corresponding
emission episode. Second,
the observed decay is the superposition of the curvature decay and
underlying afterglow decay from the external shock. One needs to
subtract the underlying afterglow contribution before performing the
test. The credibility of the curvature effect interpretation to the
steep decay phase is that by properly taking into account the two
effects mentioned above, the steep decay is consistent
with Eq.(\ref{curvature}) with $t_0$ shifted to the beginning of the
last pulse of prompt emission (Liang et al. 2006) at least in some
cases.

Besides the standard curvature effect model, other 
interpretations for the steep decay phase 
have been discussed in the literature. 
\begin{itemize}
\item In some cases, the steep-decay slope may be shallower than 
the expectation of the curvature effect\footnote{It is worth noticing
that generally a decay slope steeper than the curvature effect prediction
is not allowed, unless the jet is very narrow. Usually, even if
the intrinsic temporal decay slope is steeper than Eq.(\ref{curvature}),
the curvature effect nonetheless takes over to define the decay slope.}.
This would suggest that the central engine may not die abruptly or the
shocked region may not cool abruptly, but
rather decay with time gradually, leading to a decaying 
afterglow related to the central engine 
(Fan \& Wei 2005; Zhang et al. 2006). This was taken
by Fan et al. (2006) to interpret the abnormal power law decay of
GRB 060218 (Campana et al. 2006b).
\item Yamazaki et al. (2006) study the curvature effect of an
inhomogeneous fireball (mini-jets). They found that the decay tail
is generally smooth, but sometimes could have structures, which may
interpret the small-scale structure in some of the decay tails.
\item Pe'er et al. (2006) suggest that the emission from the
relativistically expanding hot plasma ``cocoon'' associated with 
the GRB jet could also give rise to the steep decay phase
observed by Swift.
\end{itemize}

Motivated by the discovery of the spectrally evolving tails
in GRB 050724 (Campana et al. 2006b) and GRB 060614 (Gehrels et al.
2006; Zhang et al. 2007a), recently Zhang et al. 
(2007c) performed a systematic time-dependent spectral analysis of 17 
bright steep decay tails. They found that while 7 tails show no
apparent spectral evolution, the other 10 do. A simple curvature
effect model invoking an angle-dependent spectral index cannot 
interpret the data. This suggests that the curvature effect is not 
the sole factor to control the steep decay tail phase at least in 
some bursts. Zhang et al. (2007c) 
show that some of the spectrally evolving tails may be interpreted
as superposition of the curvature effect tail and an underlying 
central engine afterglow, which is soft but decays ``normally''.
Such a component has been seen in GRB 060218 (Campana et al. 2006b),
which cannot be interpreted by the standard external shock afterglow
(Willingale et al. 2006). The strong spectral evolutions in GRB 050724,
GRB 060218, and GRB 060614, however, cannot be interpreted with such
a model. They are interpreted as internal shock afterglows by Zhang
et al. (2007c).

\subsection{X-ray flares: restarting the central engine}
 \label{sec:flares}

The X-ray flares have the following observational properties 
(Burrows et al. 2005b; Chincarini et al. 2007): 
Rapid rise and fall times with $\delta t/t_{peak}
\ll 1$; many light curves have evidence for a same decaying afterglow
component before and after the flare; multiple flares are observed
in some bursts with similar properties; large flux increases at the
flares; typically degrading fluence of flares with time, but in rare
cases (e.g. GRB 050502B) the flare fluence could be comparable with 
that of the prompt emission; 
flares soften as they progress; and later flares are less
energetic and more broadened than early flares. These properties 
generally favor the interpretation that most of them are not associated 
with external-shock related events. Rather they are the manifestations
of internal dissipations at later times, which requires restarting
the GRB central engine (Burrows et al. 2005b; Zhang et al. 2006; 
Fan \& Wei 2005; Ioka et al. 2005; Wu et al. 2006; Falcone et al.
2006; Romano et al. 2006; Lazzati \& Perna 2006). Compared with the 
external shock related models, the late 
internal dissipation models have the following two major advantages
(Zhang et al. 2006): First, since the clock needs to be re-set each 
time when the central engine restarts, it is very natural to explain 
the very sharp rising and falling lightcurves of the flares. Second,
energetically the late internal dissipation model is very economical.
While in the refreshed external shock models a large energy budget is
needed (the injection energy has to be at least comparable to that
already in the blastwave in order to have any significant injection
signature, Zhang \& M\'esz\'aros 2002), the internal model
only demands a small fraction of the prompt emission energy to
account for the distinct flares. 

The leading candidate of the late internal dissipation model is the
late internal shock model. In such a model, the collisions could be
between the fast shells injected later and the slow shells injected
earlier during the prompt phase (e.g. Zou et al. 2006; Staff et al.
2006) or between two shells injected at later times (see Wu et al.
2006 for a categorization of different types of collisions). One
concern is whether later collisions between two slow shells injected
during the prompt phase could give rise to the observed X-ray flares.
This is generally not possible. The internal shock radius can be
expressed as $R_{is} \sim d_0/(\beta_f-\beta_s) \sim 
\Gamma_s^2 d_0$, where $d_0$ is the initial separation between the 
two colliding shells, $\beta_f$ and $\beta_s$ 
are the dimensionless velocities 
of the fast and slow shells, respectively, and $\Gamma_s$ is the 
Lorentz factor of the slow shell. The second approximation is valid
if $\Gamma_f \gg \Gamma_s$. In order to produce late
internal shocks, the two slow shells must both have a low
enough Lorentz factor so that at the time of collision they do not
collide with the decelerating blastwave. Also in order not to
collide with each other earlier, their relative Lorentz factor
$\Delta \Gamma$ must be very small. When they collide,
the internal energy [$\propto (\Gamma_{sf}-1) \ll 1$] 
is usually too small to give rise to 
significant emission (where $\Gamma_{sf} \sim (\Gamma_f/
\Gamma_s+\Gamma_s/\Gamma_f)/2$ is the relative Lorentz factor between
the two shells). Should such a collision occur, most 
likely it has no interesting observational effect (see Lazzati \&
Perna 2006 for more detailed discussion on this issue). Generally,
in the internal shock model the observed time sequence reflects the
time sequence in the central engine (Kobayashi et al. 1997). As a 
result, the observed X-ray flares $(10^2-10^5)$s after the prompt
emission must imply that the central engine restarts during this
time span, say, as late as days after the prompt emission is over.

The late internal dissipation model of X-ray flares is also tested
by Liang et al. (2006). The same logic of testing the steep decay
component is used. The starting assumption is that the decay of 
X-ray flares are controlled by the curvature effect after the
abrupt cessation of the internal dissipation, so that Eq.(\ref{curvature})
is assumed to be valid. After subtracting the underlying forward
shock afterglow contribution, Liang et al. (2006) search for the 
valid zero time points ($t_0$) for each flare to allow the decay
slope satisfying the requirement of the curvature effect model.
If the hypothesis is correct, $t_0$ should be generally before
the rising segment of each flare. The testing results are impressive: 
Most of the flares indeed have their $t_0$ at the beginning of the 
flares. This suggests that the internal dissipation model 
is robust for most of the flares. It is worth emphasizing that even 
the late slow bump at around 1 day following the short
GRB 050724 (Barthelmy et al. 2005c; Campana et al. 2006a) satisfies
the curvature effect model, suggesting that the central engine is 
still active even at 1 day after the trigger. 
This is also consistent with the
late Chandra observation of this burst (Grupe et al. 2006a) that
indicates that the afterglow resumes to the pre-flare decay slope 
after the flare.

Having identified the correct model for the flare phenomenology,
one is asked about a fundamental question: how to restart the
central engine. No central engine models in the pre-Swift era have
specifically predicted extended activities far after the prompt
emission phase. Prompted by the X-ray flare observations, the
following suggestions have been made recently, and none is proved
by robust numerical simulations at the moment.
\begin{itemize}
\item Fragmentation or gravitational instabilities in the massive
star envelopes. King et al. (2005) argued that the collapse of
a rapidly rotating stellar core leads to fragmentation. The delay
of accretion of some fragments after the major accretion lead to
X-ray flares following collapsar-related GRBs. 
\item Fragmentation or gravitational instabilities in the accretion
disk. Observations of GRB 050724 (Barthelmy et al. 2005c; Campana
et al. 2006a; Grupe et al. 2006a), a short GRB associated 
with an elliptical host galaxy that is consistent with the 
compact star merger progenitor
model, reveal that it is also followed by several X-ray flares starting
from 10s of seconds all the way to $\sim 10^5$s. The properties of
these X-ray flares are similar to those in long GRBs. The 
requirement that both long and short GRBs should produce X-ray
flares with similar properties prompted Perna et al. (2006) to
suggest that fragmentation in the accretion disk, the common
ingredient in both long and short GRB models, may be the agent for
episodic accretion that powers the flares.
\item Magnetic barrier. Based on the MHD numerical simulations in 
other contexts and theoretical arguments, Proga \& Zhang (2006) 
argue that the magnetic barrier near the black hole may act as an 
effective modulator of the accretion flow. The accretion flow can be 
intermittent in nature due to the role of magnetic fields. This
model does not require the flow being chopped (e.g. due to 
fragmentation or gravitational instabilities) at larger radii before
accretion, although in reality both processes may occur altogether.
The magnetic barrier model is in accordance with the magnetic origin
of X-ray flares based on the energetics argument (Fan et al. 2005c).
\item NS-BH merger. Flares in GRB 050724 (Barthelmy et al. 2005c)
pose great challenge to the previous compact star merger models.
Numerical simulation of NS-NS mergers typically gives a short
central engine time scale (0.01-0.1)s, if the final product is a
BH-torus system
(Aloy et al. 2005). In order to account for the late time flares
in 050724, Barthelmy et al. (2005c) suggest a possible NS-BH merger
progenitor system. Numerical simulations of BH-NS merger systems 
have been performed. Although X-ray flares at 100s of seconds 
or later still challenge the model, extended accretion over several
seconds could be reproduced (Faber et al. 2006; cf. Rosswog 2005).
\item NS-NS merger with a postmerger millisecond pulsar. Dai et al.
(2006a) argue a possible solution for the extended X-ray flares 
following merger-type GRBs. Numerical simulations have shown that
the product of a NS-NS merger may not be a BH (Shibata et al.
2005), if the NS equation-of-state is stiff. Instead, the final
product may be a differentially-rotating massive neutron star.
If the initial magnetic fields of the NS is not strong, 
the $\alpha-\Omega$ dynamo action would induce magnetic explosions 
that give rise to late internal shocks to produce 
X-ray flares (Dai et al. 2006a).
\item Multi-stage central engine. Gao \& Fan (2006) and Staff et al.
(2006) proposed multi-stage central engine models to interpret 
X-ray flares.
\end{itemize}

\subsection{Shallow decay phase: still a mystery}

The shallow decay phase could follow the steep decay phase or 
immediately follow the prompt emission (O'Brien et al. 2006;
Willingale et al. 2006). It is very likely related to the 
external-shock-origin afterglow. However,
the origin of this shallow decay phase is more difficult to identify,
since there exist several different possibilities that are not easy 
to differentiate among each other from the X-ray observations. The 
fact that the spectral index does not change across the temporal
break from the shallow decay phase to the normal decay phase 
rules out the
models that invoke crossing of a spectral break across the band.
The nature of the break should be then either hydrodynamical or 
geometrical.

Following models have been discussed in the literature.

\begin{itemize}
\item Energy injection invoking a long-term central engine. The most
straightforward interpretation of the ``shallower-than-normal'' phase
is that the total energy in the external shock continuously increases
with time. This requires substantial energy injection into the 
fireball during the phase (Zhang et al. 2006; Nousek et al. 2006; 
Panaitescu et al. 2006a). There are two possible energy injection
schemes (Zhang et al. 2006; Nousek et al. 2006). The first one is to 
simply invoke a long-lasting central
engine, with a smoothly varying luminosity, e.g. $L\propto t^{-q}$
(e.g. Zhang \& M\'esz\'aros 2001). In order to give interesting 
injection signature $q<1$ is required; otherwise the total energy 
in the blastwave essentially does not increase with time. Such a 
possibility is valid for the central engines invoking a spinning down
pulsar (Dai \& Lu 1998; Zhang \& M\'esz\'aros 2001; Fan \& Xu 2006)
or a long-lasting BH-torus system (MacFadyen et al. 2001). One
possible signature of this scenario that differentiates it from the
varying-$\Gamma$ model discussed below is that a strong relativistic
reverse shock is usually expected, if at the shock interacting region
the $\sigma$-parameter (the ratio between the Poynting flux and the
kinetic flux) is degraded to below unity (Dai 2004; Yu \& Dai 2006).
Alternatively, if $\sigma$ is still high at the shock region, 
the reverse shock may be
initially weak, but would still become relativistic if the engine
lasts long enough (i.e. this is effectively 
a rather thick shell, Zhang \& Kobayashi
2005). The observational data suggest a range of $q$ values with
typical value $q\sim 0.5$. This is different from the requirement
of the analytical pulsar model ($q=0$). However, numerical calculations
suggest that a pulsar model can fit some of the XRT lightcurves
(Fan \& Xu 2006; De Pasquale et al. 2006; Yu \& Dai 2006).
\item Energy injection from the ejecta with a wide $\Gamma$-distribution.
This model invokes a distribution of the Lorentz factor of the 
ejecta with low-$\Gamma$ ejecta lagging behind the high-$\Gamma$ ones,
and only piling up to the blastwave when the high-$\Gamma$ part is 
decelerated (Rees \& M\'esz\'aros 1998). In order to produce a 
smooth power law decay, the $\Gamma$-distribution needs to be close
to a power law with $M(>\Gamma) \propto \Gamma^{-s}$. A significant
energy injection requires $s>1$. The temporal break around $(10^3
-10^4)$ s suggests a cutoff of Lorentz factor around several 10's,
below which $s$ becomes shallower than unity (Zhang et al. 2006).
Granot \& Kumar (2006) have used this property to constrain the
ejecta Lorentz factor distribution of GRBs within the
framework of this model.
The reverse shock of this scenario is typically non-relativistic
(Sari \& M\'esz\'aros 2000), since the relative Lorentz factor
between the injection shell and the blastwave is always low
when the former piles up onto the latter.
\item Delayed energy transfer to the forward shock. Analytically,
the onset of afterglow is estimated to be around $t_{dec} =
{\rm max}(t_\gamma, T)$, where $t_\gamma \sim 5~{\rm s}
(E_{K,52}/n)^{1/3} (\Gamma_0/300)^{-8/3}(1+z)$ is the time scale
at which the fireball collects $\Gamma^{-1}$ of the rest mass of
the initial fireball from the ISM, and $T$ is the duration of the 
explosion. The so-called ``thin'' and ``thick'' shell cases
correspond to $t_\gamma > T$ and $t_\gamma < T$, respectively
(Sari \& Piran 1995; Kobayashi et al. 1999). Numerical
calculations suggest that the time scale before entering the
Blandford-McKee self-similar deceleration phase is long, of
order several $10^3$ s (Kobayashi \& Zhang 2006). This suggests
that it takes time for the kinetic energy of the fireball to
be transferred to the medium. In a high-$\sigma$ fireball, 
there is no energy transfer during the propagation of a reverse
shock (Zhang \& Kobayashi 2005). Although
energy transfer could happen after the reverse shock disappears,
this potentially further delays the energy transfer process
(although detailed numerical simulations are needed to verify this).
The shallow decay phase may simply reflect the slow energy
transfer process from the ejecta to the ambient medium. This model 
(e.g. Kobayashi \& Zhang 2006) predicts a significant curvature
of the lightcurves. This is consistent with some of the lightcurves
that show an early ``dip'' before the shallow decay phase. For those
cases with a straight shallow decay lightcurve, one needs to 
incorporate the steep decay tail to mimic the observations.
\item Off-beam jet model. Geometrically one can invoke an
off-beam jet configuration to account for the shallow decay. Eichler
\& Granot (2006) show that if the line of sight is slightly outside
the edge of the jet that generates prominent afterglow emission,
a shallow decay phase can be mimicked with the combination of the
steep decay GRB tail. Toma et al. (2006) discussed a similar model
within the framework of the patchy jet models.
\item Two-component jet model. A geometric model invoking two
jet components could also fit the shallow-decay data, since additional
free parameters are invoked (Granot et al. 2006; Jin et al. 2006).
\item Precursor model. Ioka et al. (2006) suggest that if there is
a weak precursor leading the main burst, a shallow decay phase 
can be produced as the main fireball sweeps the remnants of the
precursor.
\item Varying microphysics parameter model. One could also invoke
evolution of the microphysics shock parameters to reproduce the
shallow decay phase (Ioka et al. 2006; Fan \& Piran 2006;
Granot et al. 2006; Panaitescu et al. 2006b).
\item Dust scattering model. Shao \& Dai (2006) suggest that 
small angle scattering of X-rays by dust could also give rise
to a shallow decay phase under certain conditions.
\end{itemize}

Can different possibilities be differentiated by the more abundant data?
It seems to be a challenging task. The author is inclined to the
first three interpretations on the above list. For the two energy 
injection models, one expects different reverse shock signatures 
(i.e. relativistic reverse shock for the long-term central engine 
model and non-relativistic reverse
shock for the varying-$\Gamma$ model). This would give different radio
emission properties at early times. On the other hand, the uncertainty
of the composition of the central engine outflow (e.g. the $\sigma$
parameter) would make the reverse shock signature of the former model
more obscured. The delayed energy transfer model (the third one on
the above list) is the simplest.
If it is correct, the so-called shallow decay phase is nothing but
a manifestation of the onset of afterglow (Kobayashi \& Zhang 2006).
The peak time can be then used to estimate the bulk Lorentz factor of
the fireball (which is $\sim 100$ or less for standard parameters).
This might be the case for at least some of the bursts.

\section{Optical observations}

In the pre-Swift era, the afterglow observations were mainly carried 
out in the optical and radio bands. The late time optical/radio
observations have been focused on identifying temporal breaks in
the lightcurves, which are generally interpreted as the ``jet breaks''
(see Frail et al. 2001; Bloom et al. 2003; Ghirlanda et al. 2004;
Dai et al. 2004; Friedman \& Bloom 2005; Liang \& Zhang 2005 for
compilations of the jet break data in the pre-Swift era).
Broad-band modeling was carried out for a handful of well observed
bursts (Panaitescu \& Kumar 2001, 2002; Yost et al. 2003), and the
data are generally consistent with the standard external shock
afterglow model. In some cases, very early optical flashes have been
discovered (e.g. GRB 990123, Akerlof et al. 1999; GRB 021004,
Fox et al. 2003a; GRB 021211, Fox et al. 2003b; Li et al. 2003a), 
which are generally interpreted as emission from the reverse shock
(Sari \& Piran 1999a; M\'esz\'aros \& Rees 1999; Kobayashi \& Sari
2000; Kobayashi 2000; Wang et al. 2000; 
Fan et al. 2002; Kobayashi \& Zhang 2003a; 
Zhang et al. 2003; Wei 2003; Kumar \& Panaitescu 2003; Panaitescu 
\& Kumar 2004; Nakar \& Piran 2004). Early radio flares have been 
detected in a sample of GRBs (Frail et al. 2003), which are also 
attributed to the reverse shock emission (Sari \& Piran 1999a;
Kobayashi \& Sari 2000; Soderberg \& Ramirez-Ruiz 2003). The
expectation for Swift before the launch has been that UVOT 
would collect a good sample of early afterglow lightcurves to 
allow a detailed study of GRB reverse shocks.

\subsection{Early optical afterglows: where is the reverse shock
emission?}

In the Swift era, UVOT has been regularly collecting optical photons
$\sim 100$s after the burst triggers for most GRBs. Ground-based
robotic telescopes (e.g. ROTSE-III, PAIRITEL, RAPTOR, P60, TAROT, 
Liverpool, Faulkes, KAIT, PROMPT, etc) 
have promptly observed most targets
whenever possible. A good list of early optical detections have
been made. However, the majority of bursts have very dim or
non-detection of optical afterglows (Roming et al. 2006a).
This suggests that in most cases the reverse shock, if any, is
not significant.

\begin{figure}
  \includegraphics[height=.28\textheight,angle=0]{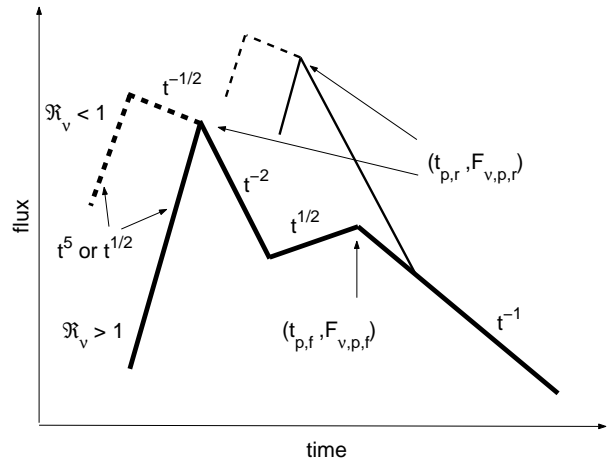}
  \caption{Theoretically expected early optical afterglow
lightcurves including emission from both reverse and forward 
shocks (from Zhang et al. 2003).}
\label{opt-lc}
\end{figure}

Figure \ref{opt-lc} displays the theoretically predicted early
optical afterglow lightcurves (Zhang et al. 2003) in the ISM
model\footnote{For wind models, see Wu et al. (2003); Kobayashi
\& Zhang (2003b); Kobayashi et al. (2004).}. The thick solid line 
shows two peaks: the first peak 
followed by $\sim t^{-2}$ decay is the reverse shock emission 
peak time, which is typically at the shock crossing
time ($t_{dec}$). The second peak followed by $\sim t^{-1}$
is the forward shock peak, which corresponds to the time 
when the typical synchrotron frequency $\nu_m$ crosses the
optical band. Depending on parameters, the forward shock peak
could be buried below the reverse shock component (the thin
solid line). One therefore has two cases of optical flashes:
Type I (rebrightening) and Type II (flattening). 

A unified study of both reverse shock and forward shock emission
suggests that Type I lightcurves should be generally expected,
if the shock microphysics parameters ($\epsilon_e$, $\epsilon_B$,
$p$, etc) are the same in both shocks. On the other hand, these
microphysics parameters may not be the same in both shocks.
In particular, if the central engine is strongly magnetized,
as is expected in several progenitor models, the outflow likely
carries a primordial magnetic field, which is likely amplified
at the shocks. It is then possible to have $R_B = 
(\epsilon_{B,r}/\epsilon_{B,f})^{1/2} \gg 1$ in some cases.
This is actually the condition to realize the Type II lightcurves
(Zhang et al. 2003). In order to interpret the bright optical 
flash and the subsequent Type II lightcurves in GRB 990123
and GRB 021211, one typically requires $R_B \sim 10$ or 
more (Fan et al. 2002; Zhang et al. 2003; Kumar \& Panaitescu
2003; Panaitescu \& Kumar 2004). 

The $\epsilon_B$ treatment is based on a purely hydrodynamical 
treatment of shocks
with magnetic fields put in by hand. Invoking a strong magnetic 
component in the reverse shock region raises the necessity to 
treat the dynamics more carefully with a dynamically important
magnetic field. Zhang \& Kobayashi (2005) studied the reverse
shock dynamics and emission for an outflow with an arbitrary
$\sigma$ parameter. They found that the most favorable case
for a bright optical flash (e.g. GRB 990123 and GRB 021211) is
$\sigma \sim 1$, i.e. the outflow contains roughly equal amount
of energy in magnetic fields and baryons. This is understandable:
For a smaller $\sigma$, the magnetic field in the reverse shock
region is smaller, and the synchrotron emission is weaker (see also
Fan et al. 2004). For a larger $\sigma$, the magnetic field is 
dynamically important, whose pressure dominates the outflow region. 
The shock becomes weak or does not exist at all (when $\sigma$
is large enough). 

The lack of bright optical flashes such as those observed in GRB 990123
and GRB 021211 is therefore not surprising. In order to have a bright
Type II flash, one needs happen to have an outflow with $\sigma \sim 
1$, while both larger and smaller $\sigma$'s would lead to not very
significant optical flashes. Even without additional suppression
effects, a non-relativistic shock with $\sigma=0$ would generally
give a reverse shock peak flux below the forward shock peak level
(Kobayashi 2000; Nakar \& Piran 2004; Zhang \& Kobayashi 2005). 
On the other extreme, a high-$\sigma$ flow would lead to very weak
reverse shock emission or no reverse shock at all (Zhang \& Kobayashi 
2005). Thus the tight early UVOT upper limits (Roming et al. 2006a) 
are not completely out of expectation.

Additional mechanisms to suppress optical flashes have been
discussed in the literature. Beloborodov (2005) argues that Compton
cooling of electrons by the prompt MeV photons may be a way to
suppress the optical flashes. Kobayashi et al. (2006) suggest that
a dominant synchrotron-self-Compton process in the reverse shock
region would suppress the synchrotron optical emission. Li et al. 
(2003b) and McMahon et al. (2006) suggest a pair-rich reverse
shock with weak optical emission.

Despite of the general disappointments, several bright optical
flashes have been detected in the Swift era, which could be generally
interpreted within the reverse/forward shock model discussed above.
The IR afterglow of GRB 041219A (Blake et al. 2005) is well modeled
by a Type I (rebrightening) lightcurve (Fan et al. 2005b). Another
Type II (flattening) lightcurve was detected from GRB 060111B
(Klotz et al. 2006). Marginal reverse shock signatures may be
present in GRB 050525A (Blustin et al. 2006; Shao \& Dai 2005),
GRB 050904 (Gendre et al. 2006; Wei et al. 2006),
GRB 060117 (Jelinek et al. 2006) and GRB 060108 (Oates et al. 2006).
Data suggest a second type of optical flashes, which tracks the
gamma-ray lightcurves (for GRB 041219A, Vestrand et al. 2005).
These optical flashes are likely related to internal shocks
(M\'esz\'aros \& Rees 1999), probably neutron rich (Fan et al.
2005b).

There are however cases that clearly show no reverse shock component
in the bright optical afterglows. GRB 061007 (Mundell et al. 2006;
Schady et al. 2006b) is such a case. Reaching a peak magnitude $<11$
(similar to 9th magnitude of GRB 990123), both the X-ray and optical
lightcurves show single power law decaying behavior from the very
beginning ($\sim 80$ s after the trigger). This suggests a strong 
external forward shock emission with enormous kinetic energy 
(Mundell et al. 2006) or a structured jet with very early jet 
break (Schady et al. 2006b). 
The reverse shock emission in this case is believed to
peak at the radio band (Mundell et al. 2006).

\subsection{Bumps and flares}

Wiggles and bumps have been observed in several pre-Swift GRB
optical afterglows (e.g. GRB 021004, Holland et al. 2003;
GRB 030329, Lipkin et al. 2004). 
Models to interpret these variabilities usually
invoke external shock related processes, such as density
fluctuation, inhomogeneous jets, refreshed shocks, or multiple
component jets (Lazzati et al. 2002; Heyl \& Perna 2003; 
Nakar et al. 2003; Berger et al. 2003a; Granot et al. 2003; Ioka
et al. 2005). Early optical lightcurves may contain neutron
decay signatures (Beloborodov 2003; Fan et al. 2005a).
Ioka et al. (2005) pointed out that some optical
fluctuations are difficult to interpret within any external shock
related schemes, and they require reactivation of the central 
engine.

That erratic X-ray flares generally require 
late central engine activities raises the question whether
some optical flashes/flares are also due to the same origin (but 
softer and even less energetic, e.g. Zhang 2005). Recent optical
afterglow observations reveal that anomalous 
optical afterglows seem to be
the norm (Stanek et al. 2006; Roming et al. 2006c). Although some
of them could be accommodated within the external shock related
models, some optical flares do show similar properties as X-ray
flares (e.g. $\delta t/t < 1$, Roming et al. 2006c), which
demands late central engine activities. For example, the
optical fluctuations detected in the short GRB 060313 optical
afterglows (Roming et al. 2006b) may be better interpreted as 
due to late central engine activities than due to density 
fluctuations (Nakar \& Granot 2006).

Efforts to model optical flares using the late internal shock model
have been carried out recently (Wei et al. 2006; Wei 2006). The
results suggest that for plausible parameters, even the traditional
reverse shock optical flashes such as those in GRB 990123, GRB 041219A
and GRB 060111B could be interpreted within the late internal shock
model.

\subsection{Optically bright vs. optically dark; optically luminous
vs. optically dim}

In the previous optical follow up observations, GRBs are generally
divided into two categories, optically bright and optically dark
ones (e.g. Jakobsson et al. 2004; Rol et al. 2005). 
The latter typically account for $\sim 50\%$ of the total 
population\footnote{Swift UVOT does not detect optical afterglows
for $\sim 67\%$ of the Swift bursts. Combining with ground-based
follow ups, the non-detection rate is $\sim 45\%$ (P. Roming, 2006,
private communication).}.
The discovery of the early optical flash of GRB 021211 (Fox et al.
2003b; Li et al. 2003) in the HETE-2 era had led to the {\em ansatz}
that as long as observations are performed early enough, most dark
bursts are not dark. This is now proven not the case (Roming et al.
2006a). Among the possible reasons of optical darkness, foreground
extinction, circumburst absorption, and high redshift are the best
candidates.

Among the optically bright GRBs, it is intriguing to discover that
there are two sub-categories, namely optically luminous and 
optically dim (Liang \& Zhang 2006; Nardini et al. 2006; Kann et al.
2006). The rest-frame lightcurves of GRBs with known redshifts
are found to follow two ``universal'' tracks. The rest-frame 10-hour
luminosities of the bursts with known redshifts 
show a clear bimodal distribution. The optically dim
bursts all appear to locate at redshifts lower than $\sim 1$ 
(Liang \& Zhang 2006). The origin of such a clear dichotomy is 
unknown, but is likely related to different total explosion energy 
involved in the two groups of bursts. 

\section{Global properties}

Combining the broad-band afterglow properties for different types 
of GRBs, one can peer into some global properties of GRB
afterglows.

\subsection{GRB radiative efficiency}

One interesting question is the GRB radiative efficiency, which
is defined as $\eta=E_\gamma/(E_\gamma+E_K)$, where $E_\gamma$ and
$E_K$ are isotropic gamma-ray energy and kinetic energy of the
afterglow, respectively. The reason why $\eta$ is important to
understand explosion mechanism is that it is related to the 
energy dissipation mechanism of the prompt emission, which is not
identified. The standard picture is internal shock dissipation,
which typically predicts several percent radiative efficiency
(Kumar 1999; Panaitescu et al. 1999, cf. Beloborodov 2000,
Kobayashi \& Sari 2001). Other mechanisms (e.g. magnetic dissipation)
may have higher efficiencies although detailed prediction is not
available. It is of great interest to estimate $\eta$ from the data,
which can potentially shed light onto the unknown energy
dissipation process.

In order to estimate $\eta$, reliable measurements of both $E_\gamma$
and $E_K$ are needed. While $E_\gamma$ could be directly measured
from the gamma-ray fluence if the GRB redshift is known, the 
measurement of $E_K$ is not trivial, which requires detailed 
afterglow modeling. In the pre-Swift era, attempts to estimate
$E_K$ and $\eta$ using late time afterglow data have been made
(e.g. Panaitescu \& Kumar 2001, 2002; Freedman \& Waxman 2001;
Berger et al. 2003b; Lloyd-Ronning \& Zhang 2004). The Swift XRT
observations suggest a substantial shallow decay phase in a good
fraction of GRBs (Fig.1). If this is due to energy injection, then
$E_K$ is a function of time. The $\eta$ values measured using the 
late time data are no longer reliable. For a constant energy
fireball, ideally early afterglows may be used to study radiative
loss of the fireball. However the shallow decay phase due to energy
injection smears the possible signature and makes such a diagnosis
difficult.

A systematic analysis of GRB radiative efficiencies using the first
hand Swift data is carried out by Zhang et al. (2007b). Similar analyses
using second-hand data for smaller samples of bursts were carried
out by Fan \& Piran (2006) and Granot et al. (2006). The conclusions
emerging from these studies suggest that in most cases the 
efficiency is very high (e.g. $>90\%$) if $E_K$ right after the 
burst is adopted. However, using $E_K$ at a later time 
when the injection process is
over one typically gets $\eta \sim$ several percent. The nature
of the shallow decay phase is therefore essential to understand
the efficiency. For example, if the shallow decay phase is due to
continuous energy injection, the GRB radiative efficiency must be
very high - causing problems to the internal shock model. If, however,
the shallow decay is simply due to the delay of energy transfer into the
forward shock, the GRB radiative efficiency is just the right one
expected from the internal shock model. One interesting finding of 
Zhang et al. (2007b) is that X-ray flashes may not be intrinsically
less efficient GRBs, as was expected in the pre-Swift era 
(Soderberg et al. 2004; Lloyd-Ronning \& Zhang 2004). Analyses show
that at the early deceleration time, XRFs are as efficient as 
harder GRBs (see also Schady et al. 2006a).

One of major breakthroughs made by Swift is the discoveries of the 
afterglows of short-duration GRBs and identifications of their
host galaxies (Gehrels et al. 2005; Fox et al. 2005; Villasenor
et al. 2005; Hjorth et al. 2005; Barthelmy et al. 2005c; Berger 
et al. 2005). These observations suggest that short GRBs likely 
have distinct progenitor systems, which are consistent with compact
star (NS-NS, NS-BH, etc) mergers. As far as the radiative efficiency 
is concerned, on the other hand, short GRBs are rather similar 
to long GRBs (Zhang et al. 2007b; see also Bloom et al. 2006 and
Lee et al. 2005). This suggests that both types of GRBs share the 
same radiation physics.

\subsection{Where are the jet breaks?}

If GRB outflows are collimated into the typical jet angle $\theta_j$,
an achromatic afterglow steepening break should be observed
in all energy bands at the time when the bulk Lorentz factor of the 
jet satisfies $\Gamma^{-1} = \theta_j$ (Rhoads 1999; Sari et al. 1999).
This time is called jet break time $t_j$. 

Identifying GRB jet breaks in the afterglow lightcurves is essential
to understand the geometric configuration and the 
total energetics of the
jets. In the pre-Swift era, a list of ``jet breaks'' have been
``identified'' in the optical (sometimes X-ray and radio) afterglows
(see Frail et al. 2001; Bloom et al. 2003; Ghirlanda et al. 2004;
Friedman \& Bloom 2005; Liang \& Zhang 2005 for compilations of the
jet break data). We use quotation marks here since the ``smoking-gun''
feature of the jet breaks, i.e. the achromatic behavior, was not robustly
established in any of these bursts. The best case was GRB 990510
(Harrison et al. 1999), in which clear multi-color optical breaks were 
discovered, which are consistent with being achromatic. The radio data
are also consistent with having a break around the same time.
However, based on radio data alone, one cannot robustly fit a break
time that is consistent with the optical break time (D. Frail, 2006,
private communication). Most of other previous jet breaks were claimed
using one-band data only, mostly in optical, and sometimes in X-ray or
radio. With these ``jet break times'' $t_j$, several empirical
relations have been discussed in the literature.
\begin{itemize}
\item Frail relation: Frail et al. (2001) found that the 
beaming-corrected gamma-ray energy is essentially constant, 
i.e. $E_{\gamma,iso} 
\theta_j^2 = E_j \sim$ const. Since the standard jet model predicts
$t_j \propto E_{\gamma,iso}^{1/3} \theta_j^{8/3}$ (Sari et al. 1999),
this relation is generally consistent with $E_{\gamma,iso} \propto 
t_{j}^{-1}$.
\item Ghirlanda relation: Ghirlanda et al. (2004) found that the
beaming-corrected gamma-ray energy is not constant, but is related to
the rest-frame spectral peak energy ($E_p$) through $E_p \propto
E_{\gamma,j}^{2/3}$. Again expressing $E_{\gamma,j}$ 
in terms of $E_{\gamma,iso}$
and $t_j$, this relation is effectively $E_p \propto E_{\gamma,iso}^{1/2}
t_j^{1/2}$. Notice that the Ghirlanda relation and the Frail relation
are incompatible with each other.
\item Liang-Zhang relation: Liang \& Zhang (2005) took one step back.
They discard the jet model, and only pursue an empirical relation
among three observables, namely $E_p$, $E_{\gamma,iso}$ and the
{\em optical band break time} $t_b$. The relation gives $E_p \propto
E_{\gamma,iso}^{0.52} t_b^{0.64}$. It is evident that if $t_b$ is 
interpreted as the jet break time, the Liang-Zhang relation is rather
similar to the Ghirlanda relation. However, the former has the 
flexibility of invoking chromatic temporal breaks across different
bands. So violating the Ghirlanda relation in other wavelengths
(e.g. in the X-ray band, Sato et al. 2006) 
does not necessarily disfavor the Liang-Zhang relation.
\end{itemize}

It has been highly expected that the multi-wavelength 
observatory Swift would clearly detect achromatic breaks 
in some GRBs to verify
the long-invoked GRB jet scenario. The results are however
discouraging. After detecting nearly 200 bursts, no ``textbook''
version jet break is yet detected in any GRB. The lack of 
detections may be attributed partially to the intrinsic 
faintness of the Swift afterglows, and partially to the very low 
rate of late time optical follow-up observations.
Achromatic breaks were indeed observed in some bursts, but none
satisfy the salient features expected in the jet model. For example,
GRB 050801 (Rykoff et al. 2006) and GRB 060729 (Grupe et al. 2006b)
have an early achromatic break covering both the X-ray and optical
bands. However, the break is the transition from the shallow decay
phase to the normal decay phase, which is likely an injection break
rather than a jet break. GRB 050525A (Blustin et al. 2006) has
an achromatic break in X-ray and optical bands, which might be
interpreted as a jet break. However, the post-break temporal indices
in both X-ray and optical bands are too shallow to comply with 
the $\propto t^{-p}$ prediction. An achromatic jet break was 
claimed for GRB 060526 (Dai et al. 2006b). However, the post
break indices for both X-ray and optical bands are significantly
different from each other, so that more complicated jet models 
are needed to accommodate the data.

In most other cases, data seem to disfavor (or at least not to
support) the existence of jet breaks. The data also cast doubts
on the previous identified jet breaks. These pieces of evidence
are collected in the following.
\begin{itemize}
\item Optical follow up of GRB 060206 reveals a clear temporal
break that would be regarded as a typical jet break should the X-ray
have not been collected (Monfardini et al. 2006). However, X-ray data 
show a remarkable single power law decay without any evidence of a 
break at the optical break time (Burrows 2006). 
\item Many other X-ray afterglows also show remarkable single
power law decays extending to very late times (10 days or later,
Burrows 2006).
The lower limits of the beaming-corrected gamma-ray energy
of many bursts already greatly exceed the standard energy 
reservoir value suggested by Frail et al. (2001) and Bloom et 
al. (2003) (Burrows 2006).
\item Based on the Ghirlanda relation, Sato et al. (2006) have 
searched for expected jet breaks of three Swift bursts in the 
X-ray band with null results. This suggests that Ghirlanda
relation is not a common relation satisfied by most bursts.
This fact however does not disfavor the Liang-Zhang relation,
since an optical break may still exist at the expected time
if the breaks are chromatic. Late time optical observations
are needed to test whether the Liang-Zhang relation is 
generally valid/violated for most bursts.
\end{itemize}

It is worth mentioning that in several cases, the X-ray data
are consistent with having a jet break. These include
GRBs 050315, 050814, 050820A, 051221A and 060428A (see Burrows
2006 for a review). In particular, late Chandra ToO observations
of the short GRB 051221A reveal a possible jet break, 
suggesting collimation in merger type GRBs (Burrows et al.
2006).

\subsection{A new paradigm of temporal breaks?}

The data seem to suggest that there might exist other types 
of temporal breaks at least for some bursts 
that are not related to jet breaks. 
A very interesting feature of the afterglow breaks is that
the X-ray breaks systematically lead the optical breaks,
which in turn systematically lead the radio breaks. This fact,
along with the chromatic breaks in both X-rays (e.g. 
Panaitescu et al. 2006b) and optical (e.g. Monfardini et al.
2006), drives the author to speculate an ad hoc scenario
to interpret these temporal breaks as well as the Liang-Zhang
($E_{\gamma,iso}-E_p-t_b$) relation. In this scenario, the
spectral break in the prompt gamma-ray emission ($E_p$) and
the chromatic temporal breaks in the afterglow lightcurves
may be all related to a same electron energy distribution
break that rolls down from high energy to low energy. 
Initially the break is in the 
gamma-ray band, which defines the $E_p$ in the prompt emission
spectrum. Later this break moves to the X-ray band in $\sim 
(10^3-10^4)$ s, giving rise to the early injection-like breaks
in some bursts. The break keeps moving down to the optical band
around a day, which can account for the pre-Swift optical
breaks that were interpreted as jet breaks. Later it moves
to the radio band in $\sim 10$ days. Such a scenario gives
a natural link between $E_p$ and the optical break time $t_b$
in the Liang-Zhang relation, which is otherwise difficult to
explain.

There are several problems with this scenario, however. 
First, it requires
that for the bursts of interests the prompt emission and the 
afterglows are from the same emission component. This is in
contradiction with the Swift finding that prompt emission and
X-ray flares are of the internal origin while afterglows are of 
external origin (e.g. Zhang et al. 2006). Nonetheless, maybe
some bursts indeed satisfy this requirement. If so, either the
prompt emission of these bursts are of external origin, or
more possibly, the afterglows of these bursts are of internal
origin, i.e. they are the central engine afterglows. Second,
a natural expectation of this scenario is that the spectral
indices before and after the breaks should be different. This
seems to contradict with the X-ray data that suggest no spectral
evolution across the early break is observed. However, the
latest systematic study of temporal breaks (Willingale et al.
2006) suggest that spectral changes in some breaks are 
observed. It is interesting to look closer whether those 
bursts with spectral evolution are also consistent with
both prompt emission and afterglow being from the same
emission component, and therefore might satisfy the 
temporal break scenario suggested here.

Other than these difficulties, this new scenario seems to be
able to account for chromatic breaks observed in some afterglows.
A hard test of this scenario is to find some bursts that have a break
crossing through the X-ray, optical and radio bands in turn.
Although no clear example is available in the 
Swift data sample, the previous GRB 030329 may satisfy the
requirement of this model. It has been claimed that there are
two ``jet breaks'' in this burst (Berger et al. 2003a): an early
optical break and a later radio break. These two breaks were
used to argue a two-component jet model for this burst.
Within the scenario proposed here, the two breaks are simply
the same break rolling over the optical and radio bands at
different times.

Similar to the Liang-Zhang relation that connects $E_p$ with the
optical break time $t_{b,opt}$, this scenario also predicts a
correlation between $E_p$ and the X-ray break times $t_{b,X}$.
Such a correlation seems to have been revealed in the Swift data
(Willingale et al. 2006).

\section{Conclusions: a global picture of GRB afterglows}
 \label{sec:conclusion}

Swift opens a new window to study the global properties of GRB 
afterglows. Although some model predictions are verified by the
new data (it is impressive that many X-ray afterglows satisfy
the closure relations predicted in simple external shock models),
what we gain more from the Swift observations are problems that 
challenge the previous theoretical framework. It is evident that
our view about GRB afterglows is now much broader than in 
the pre-Swift era. We tentatively draw the following conclusions.

\begin{itemize}
\item What we call afterglows actually include
two distinct components: one is from the traditional external shock,
the other is from the central engine. It is evident that some 
flaring components (most X-ray flares and probably some optical 
flares) are of internal origin, marking the reactivation of the 
central engine. However, one may be driven to accept that some
of the smooth power law decay components may also reflect the
emission from the central engine. This is relevant to some XRT
lightcurves that do not satisfy any closure relation, and to some
chromatic breaks that are difficult to accommodate within the 
standard external shock models. If some of the power-law decay
lightcurves directly reflect the luminosity output of 
the central engine, one has to accept
that the GRB central engine is long lived, not only erratically (to
produce flares), but in some cases also continuously (to produce
the smooth decay component). The latter is consistent with the first
energy injection model (the refreshed shock model) 
that invokes a long-term
central engine.
\item The most puzzling question is the nature of GRB afterglow temporal
breaks. The current data suggest that some breaks are still {\em
consistent with} being jet breaks, but in a lot of other cases
conflicts present. The pre-Swift relations invoking jet 
breaks (Frail and Ghirlanda relations) are not confirmed by the
Swift data. Evidence of chromatic breaks in both X-ray and 
optical bands is accumulating. At least for some bursts, one may
require new interpretations (such as the one proposed in \S4.3)
for the observed temporal breaks. Since the nature of breaks is one
of the major puzzles in the concurrent GRB study, late time optical
follow up observations are strongly encouraged to reveal whether
optical breaks as predicted by the $E_p-E_{\gamma,iso}-t_{b,opt}$
(Liang-Zhang) relation still exist despite of the 
apparent lack of X-ray breaks at the same epoch. This would make
a strong case for whether the previous ``jet breaks'' are in fact 
chromatic breaks.
\item It is now high time to perform systematic data analyses of the
abundant Swift GRB data to peer into the global properties of the
bursts. While one can still gain knowledge from special individual
events (such as GRB 060218 and GRB 060614), for most of the ``normal''
bursts, only global statistical properties can serve to improve our
understanding of GRBs. Issues need to clarify include the properties
and nature of broadband flares/bumps, temporal breaks, the 
compatibility of data with closure relations, etc. Some extensive
efforts in these directions have just commenced (e.g. Willingale et al.
2006; Chincarini et al. 2006).
\end{itemize}

\medskip
The author thanks Shiho Kobayashi, Paul O'Brien, Dick Willingale and
Dave Burrows for helpful commments on the paper.
This work is supported by NASA under grants NNG06GH62G and NNG05GB67G.

\section{References}


\setlength{\parskip}{.00in}
\reference
Akerlof, C. et al. Observation of contemporaneous optical radiation 
from a gamma-ray burst. \nat\vol{398}{400-402} (1999).
\reference
Aloy, M. A., Janka, H.-T. and Mller, E. Relativistic outflows from 
remnants of compact object mergers and their viability for short 
gamma-ray bursts. \aa\vol{436}{273-311} (2005).
\reference
Barthelmy, S. D. et al. The Burst Alert Telescope (BAT) on the 
SWIFT Mission, Space Science Reviews, \vol{120}{143-164} (2005a).
\reference
Barthelmy, S. D. et al. Discovery of an Afterglow Extension of the 
Prompt Phase of Two Gamma-Ray Bursts Observed by Swift, 
\apj\vol{635}{L133-L136} (2005b)
\reference
Barthelmy, S. D. et al. An origin for short gamma-ray bursts 
unassociated with current star formation, \nat\vol{438}{994-996} 
(2005c)
\reference
Beloborodov, A. M. On the Efficiency of Internal Shocks in Gamma-Ray 
Bursts. \apj\vol{539}{L25-L28} (2000).
\reference
Beloborodov, A. M. Neutron-fed Afterglows of Gamma-Ray Bursts.
\apj\vol{585}{L19-L22} (2003).
\reference
Beloborodov, A. M. Optical and GeV-TeV Flashes from Gamma-Ray Bursts.
\apj\vol{618}{L13-L16} (2005).
\reference
Berger, E. et al. A common origin for cosmic explosions inferred from 
calorimetry of GRB030329. \nat\vol{426}{154-157} (2003a).
\reference
Berger, E. Kulkarni, S. R. and Frail, D. A. A Standard Kinetic Energy 
Reservoir in Gamma-Ray Burst Afterglows. \apj\vol{590}{379-385} (2003b).
\reference
Berger, E. et al. The afterglow and elliptical host galaxy of the 
short gamma-ray burst GRB 050724. \nat\vol{438}{988-990} (2005).
\reference
Blake, C. H. et al. An infrared flash contemporaneous with the 
gamma-rays of GRB 041219a. \nat\vol{435}{181-184} (2005).
\reference
Blandford, R. D. and McKee, C. F. Fluid dynamics of relativistic 
blast waves. Physics of Fluids, \vol{19}{1130-1138} (1976)
\reference
Bloom, J. S. Frail, D. A., and Kulkarni, S. R. Gamma-Ray Burst Energetics 
and the Gamma-Ray Burst Hubble Diagram: Promises and Limitations.
\apj\vol{594}{674-683} (2003).
\reference
Bloom, J. S. et al. Closing in on a Short-Hard Burst Progenitor: 
Constraints from Early-Time Optical Imaging and Spectroscopy of a 
Possible Host Galaxy of GRB 050509b. \apj\vol{638}{354-368} (2006).
\reference
Blustin, A. J. et al. Swift Panchromatic Observations of the Bright 
Gamma-Ray Burst GRB 050525a. \apj\vol{637}{901-913} (2005).
\reference
Burrows, D. N. Swift X-ray afterglows: where are the X-ray jet breaks?
Nuovo Cimento, submitted (2006).
\reference
Burrows, D. N. et al. The Swift X-Ray Telescope, Space Science Reviews, 
\vol{120}{165-195} (2005a).
\reference
Burrows, D. N. et al. Bright X-ray Flares in Gamma-Ray Burst Afterglows, 
Science, \vol{309}{1833-1835} (2005b).
\reference
Burrows, D. N. et al. Jet Breaks in Short Gamma-Ray Bursts. II: The 
Collimated Afterglow of GRB 051221A. \apj\vol{653}{468-473} (2006).
\reference
Campana, S. et al. Swift Observations of GRB 050128: The Early X-Ray 
Afterglow. \apj\vol{625}{L23-L26} (2005).
\reference
Campana, S. et al. The X-ray afterglow of the short gamma ray burst 
050724, \aa\vol{454}{113-117} (2006a).
\reference
Campana, S. et al. The association of GRB 060218 with a supernova and 
the evolution of the shock wave, \nat{442}{1008-1010} (2006b).
\reference
Chevalier, R. A. and Li, Z. Y.  Wind Interaction Models for Gamma-Ray 
Burst Afterglows: The Case for Two Types of Progenitors.
\apj\vol{536}{195-212} (2000).
\reference
Chincarini, G. et al. Prompt and afterglow early X-ray phases in the 
comoving frame. Evidence for Universal properties? astro-ph/0506453
(2005).
\reference
Chincarini, G. et al. Flares in gamma-ray bursts: I. morphology and
timing properties as observed by Swift. in preparation (2006).
\reference
Cusumano, G. et al. Swift XRT Observations of the Afterglow of GRB 
050319. \apj\vol{639}{316-322} (2006).
\reference
Dai, X. et al. Optical and X-Ray Observations of GRB 060526: A Complex 
Afterglow with An Achromatic Jet Break. \apj, submitted
(astro-ph/0609269) (2006).
\reference
Dai, Z. G. Relativistic Wind Bubbles and Afterglow Signatures.
\apj\vol{606}{1000-1005} (1998).
\reference
Dai, Z. G., Liang, E. W., and Xu, D. Constraining $\Omega_M$ and Dark 
Energy with Gamma-Ray Bursts. \apj\vol{612}{L101-L104} (2004).
\reference
Dai, Z. G. and Lu, T. gamma-Ray Bursts and Afterglows from Rotating 
Strange Stars and Neutron Stars, \prl\vol{81}{4301-4304} (1998).
\reference
Dai, Z. G., Wang, X. Y., Wu, X. F. and Zhang, B. X-ray Flares from 
Postmerger Millisecond Pulsars, Science, \vol{311}{1127-1129} (2006).
\reference
De Pasquale, M. et al. Swift observations of GRB 050712,
\mnras\vol{370}{1859-1866} (2006).
\reference
Dyks, J., Zhang, B. and Fan, Y. Z. Curvature effect in structured GRB 
jets. preprint, astro-ph/0511699 (2006).
\reference
Eichler, D.,  Granot, J. The Case for Anisotropic Afterglow Efficiency 
within Gamma-Ray Burst Jets \apj\vol{641}{L5-L8} (2006).
\reference
Faber, J. A. et al. General Relativistic Binary Merger Simulations and 
Short Gamma-Ray Bursts \apj\vol{641}{L93-L96} (2006).
\reference
Falcone, A. D. et al. The Giant X-Ray Flare of GRB 050502B: Evidence 
for Late-Time Internal Engine Activity, \apj\vol{641}{1010-1017}
(2006).
\reference
Fan, Y.-Z., Dai, Z.-G., Huang, Y.-F. and Lu, T. Optical Flash of GRB 
990123: Constraints on the Physical Parameters of the Reverse Shock.
Ch.J.A.A. \vol{2}{449-453} (2002).
\reference
Fan, Y. Z. and Piran, T. Gamma-ray burst efficiency and possible 
physical processes shaping the early afterglow. \mnras\vol{369}
{197-206} (2006).
\reference
Fan, Y. Z., Piran, T. and Xu, D. The interpretation and implication of 
the afterglow of GRB 060218. JCAP, \vol{09}{013} (2006).
\reference
Fan, Y. Z. and Wei, D. M. Late internal-shock model for bright X-ray 
flares in gamma-ray burst afterglows and GRB 011121.
\mnras\vol{364}{L42-L46} (2005).
\reference
Fan, Y. Z., Wei, D. M., and Wang, C. F. The very early afterglow 
powered by ultra-relativistic mildly magnetized outflows.
\aa\vol{424}{477-484} (2004).
\reference
Fan, Y. Z. and Xu, D. The X-ray afterglow flat segment in short 
GRB 051221A: energy injection from a millisecond magnetar?
\mnras\vol{372}{L19-L22} (2006).
\reference
Fan, Y. Z., Zhang, B. and Wei, D. M. Early optical afterglow 
lightcurves of neutron-fed gamma-ray bursts. \apj\vol{628}{298-314}
(2005a).
\reference
Fan, Y. Z., Zhang, B. and Wei, D. M. Early Optical-Infrared Emission 
from GRB 041219a: Neutron-rich Internal Shocks and a Mildly Magnetized 
External Reverse Shock. \apj\vol{628}{L25-L28} (2005b).
\reference
Fan, Y. Z., Zhang, B. and Proga, D. Linearly Polarized X-Ray Flares 
following Short Gamma-Ray Bursts. \apj\vol{635}{L129-L132} (2005c).
\reference
Fenimore, E. E.,  Madras, C. D. and  Nayakshin, S. Expanding Relativistic 
Shells and Gamma-Ray Burst Temporal Structure, \apj\vol{473}{998-1012}
(1996).
\reference
Fox, D. W. et al. Early optical emission from the gamma-ray burst of 
4 October 2002. \nat\vol{422}{284-286} (2003a)
\reference
Fox, D. W. et al. Discovery of Early Optical Emission from GRB 021211.
\apj\vol{596}{L5-L8} (2003b)
\reference
Fox, D. B. et al. The afterglow of GRB 050709 and the nature of the 
short-hard gamma-ray bursts. \nat\vol{437}{845-850} (2005).
\reference
Frail, D. A. et al. Beaming in Gamma-Ray Bursts: Evidence for a 
Standard Energy Reservoir. \apj\vol{562}{L55-L58} (2001).
\reference
Frail, D. A. et al. A Complete Catalog of Radio Afterglows: The First 
Five Years. AJ, \vol{125}{2299-2306} (2003).
\reference
Freedman, D. L. and Waxman, E. On the Energy of Gamma-Ray Bursts.
\apj\vol{547}{922-928} (2001).
\reference
Friedman, A. S. and Bloom, J. S. Toward a More Standardized Candle 
Using Gamma-Ray Burst Energetics and Spectra. \apj\vol{627}{1-25} (2005).
\reference
Gao, W. H. and Fan, Y. Z.  Short-living Supermassive Magnetar Model 
for the Early X-ray Flares Following Short GRBs, ChJAA, \vol{6}{513-516}
(2006).
\reference
Gehrels, N. et al. The Swift Gamma-Ray Burst Mission.
\apj\vol{611}{1005-1020} (2004).
\reference
Gehrels, N. et al. A short gamma-ray burst apparently associated with 
an elliptical galaxy at redshift $z = 0.225$. \nat\vol{437}{851-854}
(2005).
\reference
Gehrels, N. et al. A new gamma-ray burst classification scheme from
GRB 060614. \nat\vol{444}{1044-1046} (2006)
\reference
Gendre, B. et al. The Gamma-ray burst 050904: evidence for a 
termination shock? \aa, in press (astro-ph/0603431) (2006).
\reference
Ghirlanda, G, Ghisellini, G., and Lazzati, D. The Collimation-corrected 
Gamma-Ray Burst Energies Correlate with the Peak Energy of Their 
$\nu F_\nu$ Spectrum. \apj\vol{616}{331-338} (2004).
\reference
Granot, J., K\"onigl, A. and Piran, T. Implications of the early X-ray 
afterglow light curves of Swift gamma-ray bursts. \mnras\vol{370}
{1946-1960} (2006).
\reference
Granot, J. and Kumar, P. Distribution of gamma-ray burst ejecta energy 
with Lorentz factor. \mnras\vol{366}{L13-L16} (2006).
\reference
Granot, J., Nakar, E. and Piran, T. Astrophysics: refreshed shocks from 
a gamma-ray burst. \nat\vol{426}{138-139} (2003).
\reference
Grupe, D. et al. Jet Breaks in Short Gamma-Ray Bursts. I: The 
Uncollimated Afterglow of GRB 050724. \apj\vol{653}{462-467}
(2006a).
\reference
Grupe, D. et al. Swift and XMM-Newton observations of the extraordinary
GRB 060729. \apj, submitted (2006b).
\reference
Harrison, F. A. et al. Optical and Radio Observations of the Afterglow 
from GRB 990510: Evidence for a Jet. \apj\vol{523}{L121-L124} (1999).
\reference
Heyl, J. S. and Perna, R. Broadband Modeling of GRB 021004.
\apj\vol{586}{L13-L16} (2003).
\reference
Hjorth, J. et al. The optical afterglow of the short gamma-ray burst GRB 
050709. \nat\vol{437}{859-861} (2005).
\reference
Holland, S. T. et al. Optical Photometry of GRB 021004: The First Month.
AJ, \vol{125}{2291-2298} (2003).
\reference
Ioka, K., Kobayashi, S. and  Zhang, B. Variabilities of Gamma-Ray Burst 
Afterglows: Long-acting Engine, Anisotropic Jet, or Many Fluctuating 
Regions? \apj\vol{631}{429-434} (2005).
\reference
Ioka, K.,  Toma, K.,  Yamazaki, R. and Nakamura, T. Efficiency crisis 
of swift gamma-ray bursts with shallow X-ray afterglows: prior activity 
or time-dependent microphysics? \aa\vol{458}{7-12} (2006).
\reference
Jakobsson, P. et al. Swift identification of dark gamma-ray bursts.
\apj\vol{617}{L21-L24} (2004).
\reference
Jelinek, M. et al. The bright optical flash from GRB 060117.
\aa\vol{454}{L119-L122} (2006).
\reference
Jin, Z. P., Yan, T., Fan, Y. Z. and Wei, D. M. The flat X-ray segment 
of GRB051221A: two component jet in short gamma-ray bursts, preprint, 
(astro-ph/0610010) (2006).
\reference
Kann, D. A., Zeh, A. and Klose, S. Signatures of Extragalactic Dust 
in pre-Swift GRB Afterglows. \apj\vol{641}{993-1009} (2006).
\reference
King, A. et al. Gamma-Ray Bursts: Restarting the Engine. 
\apj\vol{630}{L113-L116} (2005).
\reference
Klotz, A. et al. Continuous optical monitoring during the prompt 
emission of GRB 060111B. \aa\vol{451}{L39-L42} (2006).
\reference
Kobayashi, S. Light Curves of Gamma-Ray Burst Optical Flashes.
\apj\vol{545}{807-812} (2000).
\reference
Kobayashi, S., M\'esz\'aros, P. and Zhang, B. A Characteristic Dense 
Environment or Wind Signature in Prompt Gamma-Ray Burst Afterglows.
\apj\vol{601}{L13-L16} (2004).
\reference
Kobayashi, S., Piran, T. and Sari, R. Can Internal Shocks Produce the 
Variability in Gamma-Ray Bursts? \apj\vol{490}{92-98} (1997).
\reference
Kobayashi, S., Piran, T. and Sari, R. Hydrodynamics of a Relativistic 
Fireball: The Complete Evolution. \apj\vol{513}{699-678} (1999).
\reference
Kobayashi, S. and Sari, R. Optical Flashes and Radio Flares in 
Gamma-Ray Burst Afterglow: Numerical Study. \apj\vol{542}{819-828}
(2000).
\reference
Kobayashi, S. and Sari, R. Ultraefficient Internal Shocks. \apj
\vol{551}{934-939} (2001).
\reference
Kobayashi, S. and Zhang, B. GRB 021004: Reverse Shock Emission.
\apj\vol{582}{L75-L78} (2003a).
\reference
Kobayashi, S. and Zhang, B. Early Optical Afterglows from Wind-Type 
Gamma-Ray Bursts. \apj\vol{597}{455-458} (2003b).
\reference
Kobayashi, S. and Zhang, B. The Onset of Gamma-Ray Burst Afterglow.
\apj, in press (astro-ph/0608132) (2006).
\reference
Kobayashi, S., Zhang, B., M\'esz\'aros, P. and Burrows, D.
Inverse Compton X-ray Flare from GRB Reverse Shock. 
\apj, in press (astro-ph/0506157) (2006).
\reference
Kumar, P. Gamma-Ray Burst Energetics. \apj\vol{523}{L113-L116} (1999).
\reference
Kumar, P. and  Panaitescu, A. Afterglow Emission from Naked Gamma-Ray 
Bursts, \apj\vol{541}{L51-L54} (2000).
\reference
Kumar, P. and Panaitescu, A. A unified treatment of the gamma-ray 
burst 021211 and its afterglow. \mnras\vol{346}{905-914} (2003).
\reference
Lazzati, D. et al. The afterglow of GRB 021004: Surfing on density waves.
\aa\vol{396}{L5-L8} (2002).
\reference
Lazzati, D. and Begelman, M. C. Thick Fireballs and the Steep Decay in 
the Early X-Ray Afterglow of Gamma-Ray Bursts. \apj\vol{641}{972-977}
(2006).
\reference
Lazzati, D. and Perna, R. X-ray flares and the duration of engine '
activity in gamma-ray bursts, \mnras, in press (astro-ph/0610730)
(2006).
\reference
Lee, W. H., Ramirez-Ruiz, E. and Granot, J. A compact binary merger
model for the short, hard GRB 0050509B, \apj\vol{630}{L165-L168}
(2005)
\reference
Li, W. et al. The Early Light Curve of the Optical Afterglow of GRB 
021211. \apj\vol{596}{L9-L12} (2003a)
\reference
Li, Z., Dai, Z. G., Lu, T. and Song, L. M. Pair Loading in Gamma-Ray 
Burst Fireballs and Prompt Emission from Pair-rich Reverse Shocks.
\apj\vol{599}{380-386} (2003b)
\reference
Liang, E. W. and Zhang, B. Model-independent Multivariable Gamma-Ray 
Burst Luminosity Indicator and Its Possible Cosmological Implications.
\apj\vol{633}{611-623} (2005).
\reference
Liang, E. W. and Zhang, B. Identification of Two Categories of 
Optically Bright Gamma-Ray Bursts. \apj\vol{638}{L67-L70} (2006).
\reference
Liang, E. W. et al. Testing the Curvature Effect and Internal Origin 
of Gamma-Ray Burst Prompt Emissions and X-Ray Flares with Swift Data.
\apj\vol{646}{351-357} (2006).
\reference
Lipkin, Y. M. et al. The Detailed Optical Light Curve of GRB 030329.
\apj\vol{606}{381-394} (2004).
\reference
Lloyd-Ronning, N. M. and Zhang, B. On the Kinetic Energy and Radiative 
Efficiency of Gamma-Ray Bursts. \apj\vol{613}{477-483} (2004).
\reference
MacFadyen, A. I., Woosley, S. E. and Heger, A.  Supernovae, Jets, and 
Collapsars. \apj\vol{550}{410-425} (2001).
\reference
McMahon, E., Kumar, P. and Piran, T. Reverse shock emission as a probe 
of gamma-ray burst ejecta. \mnras\vol{366}{575-585} (2006).
\reference
M\'{e}sz\'{a}ros, P. Theories of Gamma-Ray Bursts. 
ARAA, \vol{40}{137-169} (2002).
\reference
M\'{e}sz\'{a}ros, P. Gamma-ray bursts, Rep. Prog. Phys., 
\vol{69}{2259-2322} (2006)
\reference
M\'{e}sz\'{a}ros, P., Rees, M. J. Relativistic fireballs and their 
impact on external matter - Models for cosmological gamma-ray bursts. 
\apj\vol{405}{278-284} (1993).
\reference
M\'{e}sz\'{a}ros, P., Rees, M. J. Optical and Long-Wavelength 
Afterglow from Gamma-Ray Bursts, \apj\vol{476}{232-237} (1997).
\reference
M\'{e}sz\'{a}ros, P.,  Rees, M. J. GRB 990123: reverse and internal 
shock flashes and late afterglow behaviour, \mnras\vol{306}{L39-L43}
(1999).
\reference
Monfardini, A. et al. High-Quality Early-Time Light Curves of GRB 060206: 
Implications for Gamma-Ray Burst Environments and Energetics.
\apj\vol{648}{1125-1131} (2006).
\reference
Mundell, C. G. et al. The Remarkable Afterglow of GRB 061007: 
Implications for Optical Flashes and GRB Fireballs. \apj, submitted
(astro-ph/0610660) (2006).
\reference
Nakar, E. and Granot, J. Smooth Light Curves from a Bumpy Ride: 
Relativistic Blast Wave Encounters a Density Jump. preprint,
(astro-ph/0606011) (2006).
\reference
Nakar, E. and Piran, T. Early afterglow emission from a reverse shock 
as a diagnostic tool for gamma-ray burst outflows. \mnras\vol{353}
{647-653} (2004).
\reference
Nakar, E., Piran, T. and Granot, J. Variability in GRB afterglows 
and GRB 021004. New Astron., \vol{8}{495-505} (2003).
\reference
Nardini, M. et al. Clustering of the optical-afterglow luminosities 
of long gamma-ray bursts. \aa\vol{451}{821-833} (2006).
\reference
Nousek, J. A. et al. Evidence for a Canonical Gamma-Ray Burst Afterglow 
Light Curve in the Swift XRT Data, \apj\vol{642}{389-400} (2006)
\reference
Oates, S. R. et al. Anatomy of a dark burst - the afterglow of GRB 060108.
\mnras\vol{372}{327-337} (2006).
\reference
O'Brien, P. T. et al. The Early X-Ray Emission from GRBs, 
\apj\vol{647}{1213-1237} (2006)
\reference
Panaitescu, A. and Kumar, P. Fundamental Physical Parameters of 
Collimated Gamma-Ray Burst Afterglows. \apj\vol{560}{L49-L52} (2001).
\reference
Panaitescu, A. and Kumar, P. Properties of Relativistic Jets in 
Gamma-Ray Burst Afterglows. \apj\vol{571}{779-789} (2002).
\reference
Panaitescu, A. and Kumar, P. Analysis of two scenarios for the early 
optical emission of the gamma-ray burst afterglows 990123 and 021211.
\mnras\vol{353}{511-522} (2004).
\reference
Panaitescu, A., M\'esz\'aros, P. et al. Analysis of the X-ray emission 
of nine Swift afterglows. \mnras\vol{366}{1357-1366} (2006a).
\reference
Panaitescu, A., M\'esz\'aros, P. et al. Evidence for chromatic X-ray 
light-curve breaks in Swift gamma-ray burst afterglows and their 
theoretical implications. \mnras\vol{369}{2059-2064} (2006b).
\reference
Panaitescu, A., Spada, M. and M\'esz\'aros, P. Power Density Spectra 
of Gamma-Ray Bursts in the Internal Shock Model. \apj\vol{522}{L105-L108}
(1999).
\reference
Pe'er, A., M\'esz\'aros, P. and Rees, M. J. Radiation from an expanding 
cocoon as an explanation of the steep decay observed in GRB early 
afterglow light curves. \apj\vol{652}{482-489} (2006).
\reference
Perna, R., Armitage, P. J. and Zhang, B., Flares in Long and Short 
Gamma-Ray Bursts: A Common Origin in a Hyperaccreting Accretion Disk, 
\apj\vol{636}{L29-L32} (2006).
\reference
Piran, T. Gamma-ray bursts and the fireball model. Phys. Rep., 
\vol{314}{575-667} (1999).
\reference
Piran, T. The physics of gamma-ray bursts, Rev. Mod. Phys., 
\vol{76}{1143-1210} (2005).
\reference
Piro, L. et al. Probing the Environment in Gamma-Ray Bursts: The Case 
of an X-Ray Precursor, Afterglow Late Onset, and Wind Versus Constant 
Density Profile in GRB 011121 and GRB 011211, \apj\vol{623}{314-324}
(2005).
\reference
Proga, D. and Zhang, B. The late time evolution of gamma-ray bursts: 
ending hyperaccretion and producing flares \mnras\vol{370}{L61-L65}
(2006).
\reference
Rees, M. J. and M\'{e}sz\'{a}ros, P. Relativistic fireballs - 
Energy conversion and time-scales. \mnras\vol{258}{41-43} (1992).
\reference
Rees, M. J. and M\'{e}sz\'{a}ros, P. Unsteady outflow models for cosmological 
gamma-ray bursts, \apj\vol{430}{L93-L96} (1994).
\reference
Rees, M. J. and M\'{e}sz\'{a}ros, P. Refreshed Shocks and Afterglow 
Longevity in Gamma-Ray Bursts \apj\vol{496}{L1-L5} (1998).
\reference
Rhoads, J. E. The Dynamics and Light Curves of Beamed Gamma-Ray Burst 
Afterglows. \apj\vol{525}{737-749} (1999).
\reference
Rol, E. et al. How special are dark gamma-ray bursts: a diagnostic
tool, \apj\vol{624}{868-879} (2005).
\reference
Romano, P. et al. X-ray flare in XRF 050406: evidence for prolonged 
engine activity, \aa\vol{450}{59-68} (2006).
\reference
Roming, P. W. A. et al. The Swift Ultra-Violet/Optical Telescope, 
Space Science Reviews,\vol{120}{95-142}(2005)
\reference
Roming, P. W. A. et al. Suppression of the Early Optical Afterglow of 
Gamma-Ray Bursts. \apj\vol{652}{1416-1422} (2006a).
\reference
Roming, P. W. A. et al. GRB 060313: A New Paradigm for Short-Hard Bursts?
\apj\vol{651}{985-993} (2006b).
\reference
Roming, P. W. A. et al. Swift UVOT observations of early optical 
afterglows. Nuovo Cimento, submitted (2006c).
\reference
Rossi, E.,  Lazzati, D. and  Rees, M. J. Afterglow light curves, 
viewing angle and the jet structure of gamma-ray bursts 
\mnras\vol{332}{945-950} (2002).
\reference
Rosswog, S. Mergers of Neutron Star-Black Hole Binaries with Small 
Mass Ratios: Nucleosynthesis, Gamma-Ray Bursts, and Electromagnetic 
Transients. \apj\vol{634}{1202-1213} (2005).
\reference
Rykoff, E. S. et al. The Anomalous Early Afterglow of GRB 050801.
\apj\vol{638}{L5-L8} (2006).
\reference
Sari, R. and  M\'{e}sz\'{a}ros, P. Impulsive and Varying Injection in 
Gamma-Ray Burst Afterglows \apj\vol{535}{L33-L36} (2000).
\reference
Sari, R. and  Piran, T. Hydrodynamic Timescales and Temporal Structure 
of Gamma-Ray Bursts. \apj\vol{455}{L143-L146}(1995)
\reference
Sari, R. and  Piran, T. GRB 990123: The Optical Flash and the Fireball 
Model, \apj\vol{517}{L109-L112} (1999a)
\reference
Sari, R. and  Piran, T. Predictions for the Very Early Afterglow and 
the Optical Flash, \apj\vol{520}{641-649}(1999b)
\reference
Sari, R.,  Piran, T. and  Narayan, R. Spectra and Light Curves of 
Gamma-Ray Burst Afterglows, \apj\vol{497}{L17-L20} (1998)
\reference
Sari, R.,  Piran, T. and  Halpern, J. P. Jets in Gamma-Ray Bursts, 
\apj\vol{519}{L17-L20} (1999).
\reference
Sato, G. et al. Swift discovery of gamma-ray bursts without jet break
features in their X-ray afterglows. \apj, in press (astro-ph/0611148) 
(2006).
\reference
Schady, P. et al. Swift UVOT Observations of X-Ray Flash 050406.
\apj\vol{643}{276-283} (2006a).
\reference
Schady, P. et al. Extreme properties of GRB 061007: a highly energetic
or a highly collimated burst? \mnras, submitted (astro-ph/0611081) 
(2006b).
\reference
Shao, L. and Dai, Z. G. A Reverse-Shock Model for the Early Afterglow 
of GRB 050525A. \apj\vol{633}{1027-1030} (2005).
\reference
Shao, L. and Dai, Z. G. Behaviors of X-ray dust-scattering and 
implications for X-ray afterglows of Gamma-ray bursts.
\apj, submitted (2006).
\reference
Shibata, M., Taniguchi, K. and Uryu, K. Merger of binary neutron stars 
with realistic equations of state in full general relativity.  PRD, 
\vol{71}{084021} (2005).
\reference
Soderberg, A. M. et al. A Redshift Determination for XRF 020903: 
First Spectroscopic Observations of an X-Ray Flash. \apj\vol{606}
{994-999} (2004).
\reference
Soderberg, A. M. and Ramirez-Ruiz, E. Flaring up: radio diagnostics 
of the kinematic, hydrodynamic and environmental properties of 
gamma-ray bursts. \mnras\vol{345}{854-864} (2003).
\reference
Staff, J., Ouyed, R. and Bagchi, M. A three stage model for the inner 
engine of Gamma Ray Burst: Prompt emission and early afterglow.
\apj, submitted (astro-ph/0608470) (2006).
\reference
Stanek, K. Z. et al. "Anomalous" Optical GRB Afterglows are Common: 
Two $z\sim 4$ Bursts, GRB 060206 and 060210. \apj, in press
(astro-ph/0602495) (2006).
\reference
Tagliaferri, G. et al. An unexpectedly rapid decline in the X-ray 
afterglow emission of long gamma-ray bursts \nat\vol{436}{985-988}
(2005)
\reference
Toma, K. et al.  Shallow Decay of Early X-Ray Afterglows from 
Inhomogeneous Gamma-Ray Burst Jets \apj\vol{640}{L139-L142} (2006).
\reference
Vaughan, S. et al. Swift Observations of the X-Ray-Bright GRB 050315. 
\apj\vol{638}{920-929} (2006)
\reference
Vestrand, W. T. et al. A link between prompt optical and prompt gamma-ray 
emission in gamma-ray bursts. \nat\vol{435}{178-180} (2005).
\reference
Villasenor, J. S. et al. Discovery of the short gamma-ray burst GRB 050709.
\nat\vol{437}{855-858} (2005).
\reference
Wang, X. Y., Dai, Z. G. and Lu, T. Intrinsic parameters of GRB 990123 
from its prompt optical flash and afterglow. \mnras\vol{319}{1159-1162}
(2000).
\reference
Wei, D. M. The afterglow of GRB 021211: Another case of reverse shock 
emission. \aa\vol{402}{L9-L12} (2003).
\reference
Wei, D. M. The GRB early optical flashes from internal shocks: 
application to GRB990123, GRB041219a and GRB060111b. \mnras, in press
(astro-ph/0605016) (2006).
\reference
Wei, D. M., Yan, T. and Fan, Y. Z. The Optical Flare and Afterglow Light 
Curve of GRB 050904 at Redshift $z=6.29$. \apj\vol{636}{L69-L72} (2006).
\reference
Willingale, R. et al. Testing the standard fireball model of GRBs using
late X-ray afterglows measured by Swift. \apj, submitted (astro-ph/0612031) 
(2006).
\reference
Wu, X. F., Dai, Z. G., Huang, Y. F. and Lu, T. Optical flashes and very 
early afterglows in wind environments. \mnras\vol{342}{1131-1138} (2003)
\reference
Wu, X. F., Dai, Z. G., et al. X-ray flares: late internal and late 
external shocks. \apj, submitted (astro-ph/0512555) (2006).
\reference
Yamazaki, R. et al. Tail emission of prompt gamma-ray burst jets. 
\mnras\vol{369}{311-316} (2006).
\reference
Yost, S. et al. A Study of the Afterglows of Four Gamma-Ray Bursts: 
Constraining the Explosion and Fireball Model. \apj\vol{597}{459-473}
(2003).
\reference
Yu, W. and Dai, Z. G. Relativistic eind bubbles and X-ray afterglows
of GRB 060813 and GRB 060814. \mnras, submitted (2006).
\reference
Zhang, B. Gamma-ray burst early afterglows. AIP Conf. Proc. Vol. 801,
106-113 (astro-ph/0509571) (2005).
\reference
Zhang, B., Fan, Y. Z., Dyks, J. et al. Physical processes shaping GRB 
X-ray afterglow lightcurves: theoretical implications from the Swift 
XRT observations. \apj\vol{642}{354-370} (2006).
\reference
Zhang, B., Zhang, B.-B., Liang, E. W., Gehrels, N., Burrows, D. N.
\& M\'esz\'aros, P. Making a Short Gamma-Ray Burst from a Long one: 
Implications for the Nature of GRB 060614. \apj Letters, in press 
(astro-ph/0612238) (2007a).
\reference
Zhang, B., Liang, E. W., Page, K. et al. GRB radiative efficiency 
derived from the Swift data: GRBs vs. XRFs, long vs. short.
\apj, in press (astro-ph/0607177) (2007b).
\reference
Zhang, B. and Kobayashi, S. Gamma-ray burst early afterglows: reverse 
shock emission from an arbitrarily magnetized ejecta. \apj\vol{628}
{315-334} (2005).
\reference
Zhang, B., Kobayashi, S. and M\'esz\'aros, P. Gamma-ray burst early 
optical afterglow: implications for the initial Lorentz factor and the 
central engine. \apj\vol{595}{950-954} (2003).
\reference
Zhang, B. and M\'esz\'aros, P. Gamma-ray burst afterglow with continous 
energy injection: signature of a highly-magnetized millisecond pulsar.
\apj\vol{552}{L35-L38} (2001).
\reference
Zhang, B. and M\'esz\'aros, P. Gamma-ray bursts with continuous energy 
injection and their afterglow signature. \apj\vol{566}{712-722} (2002a).
\reference
Zhang, B. and M\'esz\'aros, P. Gamma-ray bursts beaming: a universal 
configuration with a standard energy reservoir? \apj\vol{571}{876-879}
(2002b).
\reference
Zhang, B. and M\'esz\'aros, P. Gamma-ray bursts: progress, problems \& 
prospects. IJMPA, \vol{19}{2385-2472} (2004).
\reference
Zhang, B.-B., Liang, E.-W. and Zhang, B. Spectral Evolution of GRB Tails: 
Central Engine and Internal Shock Afterglows? \apj, submitted
(astro-ph/0612246) (2007c).
\reference
Zou, Y. C., Dai, Z. G. and Xu, D. Is GRB 050904 a Superlong Burst?
\apj\vol{646}{1098-1103} (2006).
\end{document}